# Enhancement of the Binding Energy of Charged Excitons in Disordered Quantum Wires

T. Otterburg, D.Y. Oberli, M.-A. Dupertuis, N. Moret, E. Pelucchi, B. Dwir, K. Leifer, E. Kapon

*Laboratory of Physics of Nanostructures, Ecole Polytechnique Fédérale (EPFL)*
*1015 Lausanne, Switzerland*

Negatively and positively charged excitons are identified in the spatially-resolved photoluminescence spectra of quantum wires. We demonstrate that charged excitons are weakly localized in disordered quantum wires. As a consequence, the enhancement of the "binding energy" of a charged exciton is caused, for a significant part, by the recoil energy transferred to the remaining charged carrier during its radiative recombination. We discover that the Coulomb correlation energy is not the sole origin of the "binding energy", in contrast to charged excitons confined in quantum dots.

Exciton complexes are formed from an exciton to which one or several electronic charges are bound by the Coulomb interaction. These complexes were originally described by Lampert[1], who distinguished between intrinsic complexes consisting of only charge carriers and extrinsic complexes containing at least one ion impurity. The photon resulting from the radiative recombination process of extrinsic complexes carries fundamental information on the nature of the chemical impurity and on final eigenstates associated with the optical transition[2]. For instance, it is well known that quantum confinement increases the binding energy of an exciton bound to a neutral acceptor[3] and the ionization energy of an impurity[4]. The direct observation of intrinsic excitonic complexes is, however, rendered more difficult by the inhomogeneous broadening of the optical transition in semiconductor nanostructures.

Enhancement of the binding energy of charged excitons is expected in quantum confined systems of low dimensionality. In semiconductor quantum dots (QDs) singly and multiply charged excitons were studied by photoluminescence spectroscopy in a single quantum dot[5]/[6]. The binding energy of a charged exciton complex is defined, theoretically, as the change of Coulomb correlation energies between the exciton and its charged complex assuming that the final state of the charged carrier is the lowest energy state[7]. In QD systems it is difficult to achieve a quantitative comparison with theoretical predictions because the Coulomb interactions depend sensitively on structural parameters that are not well known : e.g., the size and shape of the QD or the strain field in self-assembled QDs[8]/[9]. The measured Coulomb correlation energies of $X^+$ and $X^-$ cannot thus be readily compared in different QD systems. In contrast with the case of QDs, $X^+$ and $X^-$ in quantum wells (QWs) possess nearly the same correlation energy and thus similar binding energies [10]/[11]/[12]. Amidst all the quantum confined systems, semiconductor quantum wires (QWRs) provide an intermediate situation from the viewpoint of confinement and dimensionality. Theoretical studies of one-dimensional charged exciton complexes predict a larger binding energy for $X^+$ than for $X^-$ and, thus, an energy ordering of the optical transitions analogous to the one prevailing in QWs [13,14]. This prediction has not yet been investigated experimentally despite several optical studies of QWRs performed with a high spatial resolution [15].

In this letter, we report on the identification of each type of charged excitons in a quantum wire by tuning the excess charge carrier density by means of an external voltage applied on a Schottky contact. The spectral emission characteristics of the charged excitons are investigated with a submicron spatial resolution in order to circumvent the omnipresent inhomogeneous broadening of interband optical transitions in semiconductor nanostructures. We show, contrary to all previous theoretical expectations, that $X^-$ is more strongly bound than $X^+$. We explain how this surprising result originates from the weak localization of the exciton complexes. Combining theoretical analysis and experimental determination of both types of charged excitons we estimate the localization length of an exciton subjected to weak disorder in a quantum wire.

For these studies, we used QWRs embedded in a field-effect structure between a semi-transparent Schottky contact on the sample's surface and an ohmic back contact on the n-doped substrate. The samples were grown by organometallic chemical vapor deposition on patterned GaAs substrates [16]. The patterns were made by electron beam lithography and consisted of lateral arrays of V-grooves with a 15 µm pitch. We designed two modulation-doped structures containing a single GaAs QWR of different nominal widths surrounded by $Al_{0.32}Ga_{0.68}As$ barriers: the thickness of the crescent-shaped QWR, measured at its center, was either 3 nm (*thin* QWR sample) or 7 nm (*thick* QWR sample). Excess electrons were provided to the QWR by Si dopants, which were inserted in the barrier within a 5 nm (95 nm) thick layer that was separated from the QWR on its top (bottom) side by a 38 nm (105 nm) thick layer. A p-type background doping of about 3 x $10^{15}$ cm$^{-3}$ in the barrier was measured by Hall effect on a reference sample. The samples were processed into mesas covered with a Schottky contact gate and with an aluminum mask in which 25 apertures of 400 nm width were made. Photoluminescence (PL) was excited with an argon-ion laser at 514.5 nm through one of the apertures; it was dispersed with a spectrograph and detected with a charged-coupled-device array detector. Samples were mounted onto the cold-finger of a cryostat and their temperature was varied between 10 and 40 K.

Typical PL spectra of the thin QWR sample are displayed in Fig. 1 for a set of 6 voltages applied onto the Schottky gate with respect to the back contact. At a gate voltage of –2.4 V the QWR is nearly depleted of electrons and the spectrum mainly consists of a sharp emission line (full width at half maximum ~ 150 µeV) that is assigned to a localized neutral exciton (labeled X) [17]. Other weaker lines on the high energy side of it are likely related to either excited states of the localized exciton or to other excitons localized farther off the center of the aperture. At increasing gate voltage, a second sharp line emerges progressively on the low energy side of the exciton line. This line is ascribed to X$^-$, which consists of an electron bound to a neutral exciton. At a gate voltage of –2.2 V the neutral exciton line has disappeared from the PL spectrum and only the charged exciton line remains. The voltage dependence of the PL spectra demonstrates that the low energy component of the doublet is a negatively charged exciton. The excess electron is transferred from an ionized donor that is found in the doped layer on the far side of the QWR with respect to the Schottky gate. The spectral separation of the doublet is a measure of the *"binding energy"* of the electron on the exciton [18]: it corresponds exactly to the difference of Coulomb correlation energies between the exciton and the charged exciton when the recoil energy of the electron is zero. We determined a binding energy of about 4.6 meV and 2.8 meV for a given aperture, respectively for the thin and thick quantum wire samples. Such an enhanced binding energy of the $X^-$ in the thin QWR does indeed reflect on the stronger quantum confinement in the thin QWR with respect to that of the thick QWR. From a survey of the literature [19], we deduce that the $X^-$ binding energies in our QWRs exceed all the published values of the $X^-$ binding energies in GaAs/AlGaAs QWs of various thicknesses. An additional line, associated to the $X^+$, is also observed in the spectra ; we postpone its identification to a later discussion in this letter.

Let us turn instead to the analysis of the spectral line shape of the $X^-$. Theory predicts that the line shape of a charged exciton is asymmetrical when its motion is free in the unconfined direction (see Ref. [14]). This intrinsic asymmetry originates from momentum transfer from $X^-$ to the remaining electron in the emission process: it is a direct consequence of the different dispersions of the electron and of the $X^-$. In Fig. 2, we plot the $X^-$ spectral lines for three temperatures and the corresponding fitted curves assuming a Lorentzian line shape and, alternatively, a Gaussian line shape. The goodness of the Lorentzian fit for all three temperatures demonstrates that the $X^-$ peak is homogeneously broadened. We repeated this line shape analysis on PL spectra taken at lower gate voltages in the same temperature range: in all cases, we obtained the best fit with a Lorentzian line shape. We thus conclude from this fitting and from the observation of a symmetrical line shape that the charged exciton is localized below 40 K. Additionally, the Lorentzian line shape gave the best fit to the exciton peak. From these results we infer that both the exciton and the charged exciton are localized on a site along the QWR. Localization of the charged excitons is caused by the fluctuations of the confinement potential that originate from structural disorder at the interface. This interfacial disorder is known to give rise to localization of excitons in QWs[20] and in QWRs[21]. In modulation-doped heterostructures, an additional source of potential fluctuations is the random distribution of ionized donors in the barrier [22]. In strongly quantum confined systems, such as narrow QWs and the QWRs discussed here, the main source of fluctuations is given by interfacial disorder. This is attested to, e.g. in the PL spectra, by the monotonous increase of the inhomogeneous broadening as the size of the QWR is reduced [23].

In the context of localization, an important question is *whether the X and the X⁻ are localized on the same site along a wire*. The simultaneous observation of both the X and the X⁻ in the time-integrated PL spectra of Fig. 1 and the smooth transfer of intensity from the X to the X⁻ at increasing gate voltage are consistent with the localization on a single site for both species. We explain the presence of X and X⁻ in the same spectra by the temporal fluctuations of the electron occupation probability on one site over a time scale given by the radiative lifetime. As the voltage is raised, the chemical potential sweeps through a local minimum of the disorder potential in the QWR, which yields a larger occupation probability of an electron on that site and, hence, a more intense X⁻ peak. To confirm this interpretation, we studied the voltage dependence of the PL spectra acquired through a number of different apertures. When the QWR was fully depleted from excess electrons, the PL spectra were either consisting of a dominant X peak or of several peaks that could be attributed to neutral excitons localized on separate sites. At increasing gate voltage, each X peak in the PL spectra became part of a well resolved doublet. These observations corroborate our attribution of the localization of the X and the X⁻ on a single site.

By performing PL studies on many apertures, the influence of disorder on the X⁻ binding energy was investigated systematically. We found indeed that the doublet spectral separation fluctuated around an average value of 4.2 meV and 2.9 meV, respectively for the thin and thick QWR samples. The fluctuation of the binding energy of X⁻ (X⁺) is reported in Fig. 3 for a subset of eleven apertures that were all characterized by a prominent doublet in their respective PL spectra. We also display the spectral positions of the X on the same figure. It is quite meaningful that the energy fluctuations of the $X^-$ binding energy are not correlated to those of the spectral position of the X. Moreover, the spectral positions of the X line span a wide energy range of 4 meV whereas the distribution of $X^-$ binding energies is narrow : its standard deviation is 0.22 meV. We also observed the same effect in the study of the thick QWR sample ; the same small standard deviation is found in both samples. It should be emphasized that the distribution of the X spectral positions in the thick QWR is narrower, which is commensurate with the smaller inhomogeneous broadening of a thicker QWR. We further checked that the $X^-$ binding energies did not depend on the applied gate voltage. Thus, our observation of a distribution of $X^-$ ($X^+$) binding energies clearly results from structural disorder.

We now describe the evolution of the PL spectrum of the thin QWR sample when the gate voltage was decreased further. While at $V_g$ = -2.4 V the spectrum is dominated by the X peak, the PL spectra at a larger negative bias are characterized by a doublet with a new energy separation (2.9 meV). The low energy component of this doublet is assigned to the $X^+$, which is formed when excess holes are present in the QWR. The progressive intensity increase of the $X^+$ peak with respect to the X one means that the probability of having excess holes in the QWR increases. The build-up of excess holes in the QWR is made possible by the opposite motion of the photoexcited electron-hole pairs in the internal electric field of the device. Indeed, a small fraction of the incident photons is absorbed in the thick spacer layer (between the QWR and the substrate) generating holes that drift towards the surface and are captured by the QWR. The larger fraction of photons is absorbed between the Schottky gate and the QWR since the QWR depth below the surface (~ 290 nm) exceeds the absorption length of the photons in the barrier (~ 190 nm). Hence, a greater number of electrons can be potentially captured by the QWR when drifting away from the surface. We observed that, at each level of the laser intensity, there was an applied voltage for which the X peak dominated the emission spectrum. With increasing laser intensity the $X^+$ peak progressively gained in intensity with respect to the X peak. The implication of this behavior is that the number of holes being captured by the QWR is larger than the number of electrons. The probable mechanism is a saturation of the trap centers in the barrier, which then results in a larger fraction of holes drifting towards the QWR. Finally, we analyzed the effect of disorder on the $X^+$ binding energy by measuring the same eleven apertures as before: the binding energy was distributed around an average value of 2.9 meV with a standard deviation of 0.12 meV (see Fig. 3).

The smaller binding energy of the $X^+$ is a surprising result. To understand it, we performed a realistic numerical calculation of the charged exciton binding energies in our QWR structures. First, we define the two-dimensional confinement potential from dark-field transmission electron micrograph of the QWR cross-section. Second, we use the single-particle Hamiltonian in the effective-mass approximation for the electron and the 4 bands **k.p** Luttinger Hamiltonian for the hole for calculating the electronic band structure and extracting effective masses for electrons and holes in the unconfined direction. Third, we solve the three-particle Schrödinger equation in which we replace the exact 3D Coulomb potentials by long-range 1D effective potentials in order to account for the different quantum confinement of electrons and holes. In this way, we obtained in the case of the thin (thick) QWR a binding energy of 1.38 meV (0.94) for $X^-$ and 1.67 meV (1.36) for $X^+$. Although these values correspond to a large enhancement of the Coulomb correlation energies with respect to those in bulk GaAs, they are still much smaller than the experimentally measured "binding energies". Furthermore, the ordering of the binding energies is reversed.

In order to understand these differences, one has to consider the role of localization in our QWRs. Let us make an ansatz on a local minimum of the disorder potential : We assume that the disorder modifies the center-of-mass motion (COM) of both the X and $X^-$ ($X^+$) without altering their relative motion. The simplest local potential has the shape of a square well of finite depth[24]. With this potential, we estimated the confinement energies of the COM ($E_{COM}$) of the X, X⁻ and X⁺ and their dependence on its length, L. Localization of the COM motion leads then to an enhancement of the negatively charged exciton "binding energy" that is given by : $\Delta B(X^-) = E_{COM}(X) - E_{COM}(X^-) + E_R(e^-)$, where $E_R$ is the confinement energy (analogous to recoil energy) of the electron. A similar expression holds for X⁺. In Fig. 4, we report the dependence of the "binding energy" of each charged exciton on L. We note that the optimization of the potential parameters has a unique solution since it depends on two unknowns. The calculation yields L equal to 37 nm and a potential depth of ~ 39 meV ($V_e+V_h$) for the set of binding energies (4.0 meV, 3.0 meV). This potential length is as an effective localization length, which characterizes the effect of interface disorder on a charged exciton. It validates a posteriori Our initial assumption on the separation of the COM and relative motions is validated a posteriori because the exciton Bohr radius (about 6 nm) of the QWR is much shorter than L. Our interpretation based on localization not only explains the enhancement of the "binding energy" of both charged excitons, but also explains why the "binding energy" of the X⁻ is larger than that of the X⁺. The origin of this effect is the dominant contribution to the charged exciton binding energy of the recoil (confinement) energy of the charge carrier that remains after the radiative recombination occured.

The experimental results described in this letter exemplify a fundamental difference between charged excitons localized in a QWR and those confined in a QD: In the regime of strong confinement, the binding energy of charged excitons is solely determined by the Coulomb interaction while, in the localization regime, the kinetic energy imparted to the remaining charged carrier contributes for a large fraction to the binding energy. The optical probing of the charged exciton complexes can provide a means to quantify the effective parameters describing a local minimum of the disorder potential in weakly disordered semiconductor nanostructures. Moreover, it is complementary to techniques based either on real-space mapping of the C.O.M. wave function by near-field scanning optical microscopy [25] or on stochastic analysis by speckle interferometry of disordered semiconductor quantum wells [26].

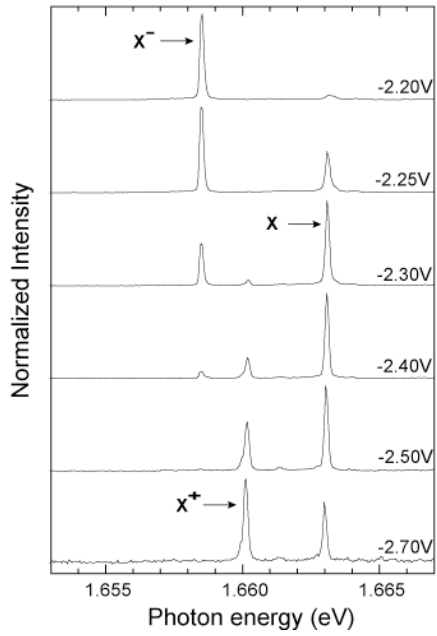

Fig. 1: Evolution of PL spectra with gate voltage for the thin QWR measured at 10 K through a submicron aperture.

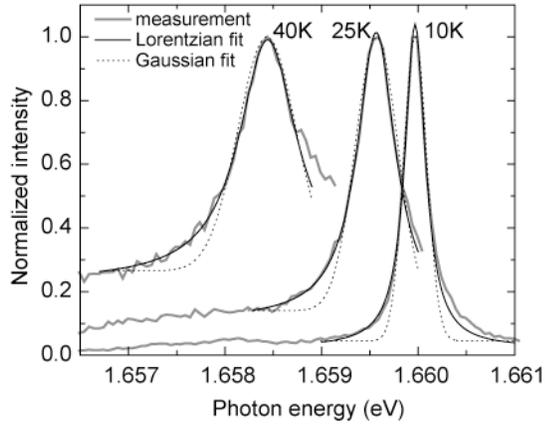

Fig. 2: Line shape fitting of the negatively charged exciton line at three temperatures.

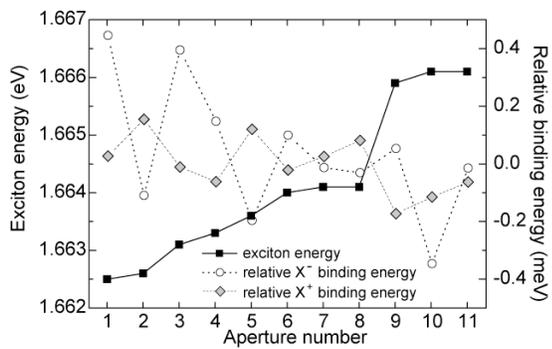

Fig. 3: Binding energies of negatively and positively charged excitons and spectral positions of neutral exciton line measured in eleven separate apertures for the thin QWR sample. The binding energies are displayed relatively to their mean value (the error is 0.05 meV).

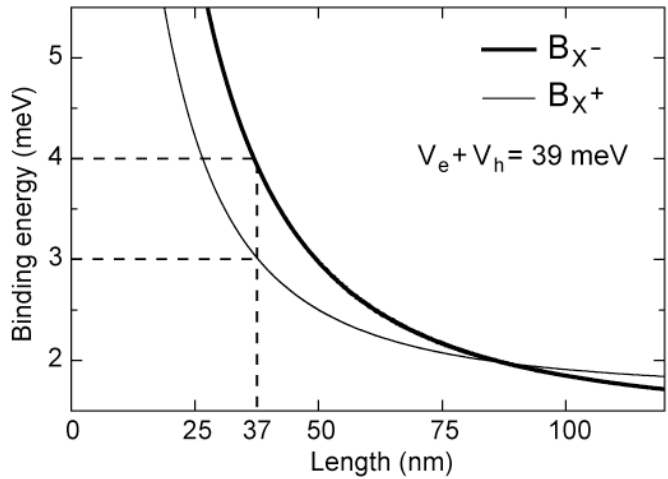

Fig. 4: Calculated dependence of the $X^+$ and $X^-$ binding energies on the length of the square potential. The depth of the potential was fitted to yield the same value of L for the binding energies of aperture No. 5.